\documentclass[twocolumn,prl,aps,superscriptaddress,showpacs]{revtex4}
\usepackage{graphicx}
\begin{document}

\title{ Calogero-Sutherland gas of ultracold Bose atoms}

\author{Yue Yu}
\affiliation{Institute of Theoretical Physics, Chinese Academy of
Sciences, P.O. Box 2735, Beijing 100080, China}
\date{\today}
\begin{abstract}
We show that the Calogero-Sutherland (C-S) gas, a famous exact
soluble one-dimensional system with an inverse square long range
interaction,  can be realized by dimension reduction in a cold
Bose atom system with a dipole-dipole interaction. Depending on
the orientation of the dipoles, the effective interaction is
either attractive or repulsive. The low-lying effective theory may
be a Luttinger liquid when the exclusion statistics parameter
$\lambda$ may be well-defined.  We hope that the C-S gas can be
realized experimentally and the Luttinger liquid character can be
observed.
\end{abstract}


 \maketitle

\noindent{\it Introduction} One-dimensional ultracold atom
systems have shown amazing physical behaviors in a bunch of
experiments in recent years. A phase transition from a
one-dimensional Bose Einstein condensate(BEC) to a Mott
insulator has been observed in a one-dimensional optical
lattice \cite{kohl}. The strong coupling limit of the
one-dimensional gas with a contact interaction, known as the
Tonks-Girardeau gas \cite{TG}, has also reached in the
ultracold Bose atoms \cite{TGE}.

Another kind of important one-dimensional systems with inverse
square interaction was known as the Calogero-Sutherland (C-S)
models \cite{ca,suth}. Nowadays the C-S models play a fundamental
role in the contemporary theoretical physics. The eigen wave
functions of these exactly soluble models may be explicitly
expressed by symmetric polynomials \cite{jack,ha}. The
quasi-particle excitations in these models obey exclusion
statistics \cite{hald1,wu,wub}. The low-lying effective theory is
a Luttinger liquid \cite{haldl,lutt} characterized by the
exclusion statistics parameter $\lambda$ \cite{wuy,ha}. It was
shown \cite{KY,wuy} that this Luttinger liquid is equivalent to a
$c=1$ conformal field theory on a circle with a radius
$R=\sqrt{2/\lambda}$ \cite{cft}. The C-S model may be a universal
Hamiltonian of some weakly disordered metals \cite{sla}. The
models may also be used to describe the edge excitations in the
fractional quantum Hall effect \cite{yu}.

 Besides mentioned above, there were already numerous
works in studying the C-S type models \cite{csmb}, but until now a
one-dimensional physical system with a real inverse square
interaction still lacks experimental realization. Recently, a
major progress in the cold atoms was that chromium atoms $^{52}$Cr
with a dipole-dipole interaction  were turned into a BEC
\cite{dpdp}.  This provides an opportunity to induce an inverse
square interaction by the dimension reduction. There already are
many theoretical considerations for the BEC with dipole-dipole
interactions in ultracold gases \cite{dpt1,dpt2}. To our
knowledge, such a dimension reduction to the dipole interaction
was not studied, which will be a main topic of this Letter.

\noindent{\it Reducing to a single-band model} With a very
strong $z$-direction confinement, a cold atom BEC cloud has
a pancake-like shape trapped in the $x$-$y$ plane. Using an
periodic optical potential along the $x$-direction, the
pancake is incised into an array of cigar-like BECs. Apply
an external field in (or perpendicular to) this $x$-$y$
plane so that the dipoles in (perpendicular to) this plane
are parallel. (See Fig. 1(b) or (c). We first neglect the
slow varying harmonic trap in the $x$-direction.) This is a
problem of the Bose atoms loaded into an optical lattice
with high filling factor. For the alkali metal atoms, the
system is described by multi-band Bose Hubbard model but can
be mapped into an effective single-band model \cite{van}. If
the product of the average number $N_0$ in a BEC and the
on-site repulsion $U$ is much larger than the trapping
potential along the $y$-direction while it is much smaller
than the strength of the optical lattice, the wave function
may be approximated by the product of the single atom ground
state wave function in the $x$-direction and the BEC's in
the $y$ direction. In this approximation, the atomic field
operator may be written as \cite{van}
\begin{eqnarray}
\psi({\bf r})\approx \sum_i a_i w(x-x_i)\Psi_{TF}(y)
\end{eqnarray}
 where
$w(x)$ is the Wannier function along the $x$-direction and
$\Psi_{TF}(y)$ is the Thomas-Fermi wave function along the
$y$-direction, which describes the ground state of atoms in
each single site. $a^\dag_i$ ($a_i$) is the creation
(annihilation) operator for an atom located in the ground
state at each site. The renormalized hopping $t_R$ and
interaction $U_R$ can be estimated in the Thomas-Fermi
approximation\cite{van}. The hopping $t$ may not be
renormalized \cite{li} while $U_R$ is reduced from its bare
value considerably due to the repulsive on-site interactions
that spread out the condensate wave function. The consistent
condition for this reduction of $U_R$ is $l_p/a\ll N_0\ll
(\hbar\omega_y/\hbar\omega_p)^2\sqrt{2\pi} L_y/a$, where $a$
is the $s$-wave scattering length, $\hbar\omega_y$ is the
trapping energy in the $y$-direction,
 $L_y=\sqrt{\hbar/m\omega_y}$ is
the effective length in the $y$-direction. $l_p$ is the
lattice spacing and $\hbar\omega_p=\hbar^2/m l_p^2$. These
conditions may easily reach in experiment.

For the atom cloud with dipole-dipole interaction, besides
the on-site interaction, there is another interaction
between two parallel dipoles located in ${\bf r}$ and ${\bf
r}'$ , which reads
\begin{eqnarray}
V_d({\bf r},{\bf r}')=d^2(1-3\cos^2\Theta)/R^3,
\end{eqnarray}
 where
$\Theta$ is the angle between ${\bf R}={\bf r}-{\bf r}'$ and
$d$ is the magnitude of the dipole moment ${\bf d}$ . For
dipoles laying in the plane, $\Theta$ is shown in Fig. 1(a)
while $\Theta=\pi/2$ for dipoles perpendicular to the plane.
The Thomas-Fermi density of dipolar BEC has the same
parabolic form as that in the short range interaction BEC.
The dipole-dipole interaction only affects the central
density and the Thomas-Fermi radius \cite{dpt1}. Therefore,
the wave function factorization approximation is also valid
if $\hbar\omega_y\ll N_0 U[1+O(\epsilon_{dd})] \ll
\hbar\omega_p$ where $U[1+O(\epsilon_{dd})]$ is the
interacting energy scale \cite{dpt1}. ($\epsilon_{dd}<1$ for
the practical system will be defined later.)

Consider a system with atom number $LN_0+N (N<L)$ where $L$
the x-direction lattice size. The effective single-band
lattice model can be obtained by keeping only the nearest
neighbor hopping and expanding on-site energy to the second
order near mean occupation $N_0$ \cite{van,li}, i.e.,
\begin{eqnarray}
H/N_0&=&-\sum_{\langle ij\rangle}
ta^\dag_ia_j+\frac{U_R}2\sum_in_i(n_i-1)\nonumber\\
&+&\sum_{i<j}U_{d,ij}n_in_j, \label{lh}
\end{eqnarray}
where $n_i=a^\dag_ia_i-N_0$; $t=\int dx
~w^*(x-x_i)(-\frac{\hbar^2}{2m}\frac{d^2}{dx^2}+V_p(x))w(x-x_j)$
for a pair of nearest neighbor sites. $U_R=\frac{4\pi\hbar^2
a}{mN_0}\int dx |w(x)|^4 \int dy |\Psi_{TF}(y)|^4$ and
$U_{d,ij}=\frac{1}{N_0} \int d{\bf r} d{\bf r}'
|w(x-x_i)|^2|w(x'-x_j)|^2V_d({\bf
r,r}')|\Psi_{TF}(y)|^2|\Psi_{TF}(y')|^2. $

This is a homogeneous model by neglecting the trapping in
the $x$-direction and describes the dynamics of the
fluctuation around the mean occupation $N_0$. Adding back
the harmonic trap with the frequency $\omega_x$, the system
has an effective size $L=\sqrt{\hbar/m\omega_x}$ in the
$x$-direction . If the width of a single BEC is wider, the
number $N_0$ of atom per site may vary site by site and
$N_0$ in the center of trapping may be several times of
$N_0$ in far away from the center. Within a cigar-like
single dipolar BEC, the dipole-dipole repulsion is strong
enough against rasing the width of the single BEC at the
trapping center. Therefore, the factorization of the ground
state wave is still valid and $N_0$ keeps an constant. The
resulting inhomogeneous model is $H$ in (\ref{lh}) added by
a term $\sum_iV_{h,i} n_i$ where
\begin{eqnarray}
V_{h,i}=\frac{1}2m\omega^2_x \int dx |w(x-x_i)|^2x^2
\end{eqnarray}
 is the
harmonic trapping potential along the $x$-direction.

\begin{figure}
\begin{center}
\includegraphics[width=8.5cm]{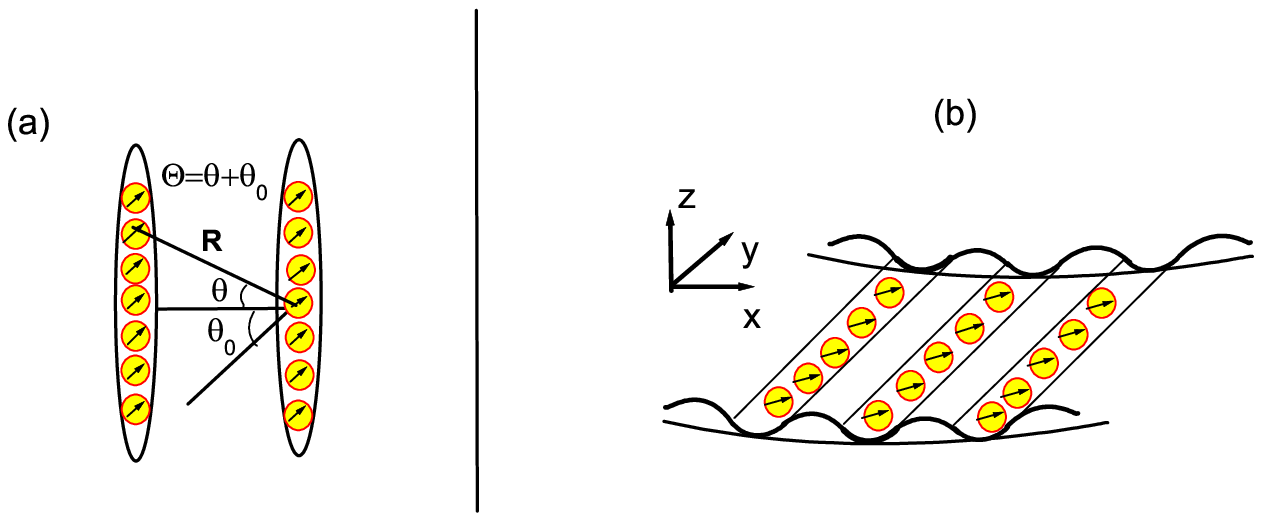}

\vspace{-1.5cm}

\includegraphics[width=4.5cm]{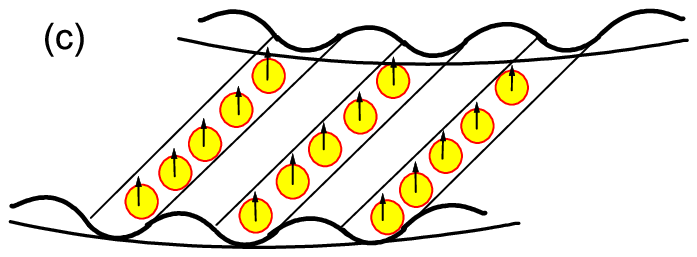}
\end{center}
 \caption{\label{fig1} (Color on-line) The trapped cold atom cloud in an optical
 lattice. (a) The
 parameters used in deriving eq. (\ref{2}) for the case (b). (b) The dipoles are in the $x$-$y$ plane.
 (c) The dipoles are perpendicular to the $x$-$y$ plane. The angle
 $\theta$ is the same as shown in (a).
 }

\vspace{-0.5cm}
\end{figure}

 We now assume
the trap in the $y$-direction is very weak such that
$|\Psi_{TF}(y)|^2$ can be approximated by the average
density $\rho_0$, which is consistent with another
approximation we will use, i.e., assuming the condensate
along the $y$-direction is infinitely long. Under these
approximations, it is easy to integrate over $y$ and $y'$
and arrives at
\begin{eqnarray}
U_{d,ij}&\approx& G_{xy(z)}\int dx dx'
|w(x-x_i)|^2|w(x'-x_j)|^2\nonumber\\&\times&\frac{1}{(x-x')^2},
\approx \frac{G_{xy(z)}}{(x_i-x_j)^2}.\label{2}
\end{eqnarray}
where $G_{xy}=-2d^2\bar\rho_0\cos^2\theta_0$ if the dipoles
orient in the plane and $G_z= 2d^2\bar\rho_0$ if dipoles is
perpendicular to the plane. Here, $\theta_0$ is the angle of
the dipole with respect to the $x$ axis (
Fig.\ref{fig1}(a)). If the on-site interaction is switched
off, to prevent  the BEC collapse due to the attractive
dipole-diploe interaction in a single BEC, one takes
$\cos^2\theta_0>\frac{1}3$ such that the dipoles in the
$y$-direction are repulsive. Thus, we have a one-dimensional
lattice model with the on-site and inverse square
interactions between the single modes from the ground state.
  If the dipoles lay in the $x$-$y$
plane (Fig. 1(b)), the effective interaction between those
atoms in different sites is attractive while it is repulsive
if the dipoles are perpendicular to the plane (Fig. 1(c)).

The stability condition of a single BEC at a given site is
given by $\epsilon_{dd}=md^2/(3\hbar^2 a)<1$ \cite{dpdp}.
For $^{52}$Cr, the experimental measured
$\epsilon_{dd}=0.159 \pm 0.034$ \cite{epsilon}. However, the
renormalized on-site interaction is reduced to  \cite{van}
\begin{eqnarray}
U_R\sim \frac{4\pi\hbar^2 a}{m}\frac{l_P}{R_{TF}}
\end{eqnarray}
 where $R_{TF}$ is the Thomas-Fermi radius of the
single BEC. On the other hand, $d$ is not renormalizable as
shown in (\ref{2}). Thus, although $\epsilon_{dd}<1$, the
renormalized on-site interaction may be much weaker than the
dipole-diploe interaction since $\epsilon_{dd}R_{TF}/l_P\gg
1$ as $\l_p/R_{TF}\ll 1$ for the cigar-like BEC we are
considered. Therefore, in the effective one-dimensional
single-band model, the dipole-dipole interaction dominates
and the on-site interaction may be turned off.

\noindent{\it Dilute gas limit and CS model}  In the dilute
gas limit with $N/L\ll1$, the $N$-particle system may be
described by a continuous model because the dispersion
$-t\cos kl_p/\hbar\sim k^2/(2m_R)$ with $m_R\sim
\hbar^2/(tl_p^2)$. Neglecting the on-site interaction as
argued above, the effective Hamiltonian reads
\begin{eqnarray}
H_{CS}=\sum_{i=1}^N\biggl(-\frac{\hbar^2}{2m_R}\frac{d^2}{dx^2_i}
+\frac{1}2m_R\omega_R^2x_i^2\biggr)+\sum_{i<
j}\frac{G_{xy(z)}}{|x_i-x_j|^2}, \label{4}
\end{eqnarray}
where the renormalized trapping potential is defined by
$m\omega_x^2=m_R\omega_R^2$. The model described by the
Hamiltonian (\ref{4}) is the famous C-S model \cite{ca,suth}
. A 'particle' at $x_i$ in this continuous model corresponds
to one more atom than $N_0$ at the lattice site $i$ in the
lattice model. The ground state wave function is given by
\begin{eqnarray}
\Psi_{0,\lambda}(x_1,...,x_N)=\prod_{ 1\leq j<k\leq N }(
x_j-x_k)^\lambda e^{-\frac{1}{2\hbar}m_R\omega_R \sum^N _jx^2_j
},\label{ground}
\end{eqnarray}
 where $\lambda$ is the solution of the equation
 $\lambda(\lambda-1)=-g_{xy}=(m_R/\hbar^2)G_{xy}$ or
 $=g_z=(m_R/\hbar^2)G_z$, respectively. The ground state energy $
E_g=\frac{1}2N\hbar\omega_R(\lambda (N-1)+1)$.

 For
$\lambda=\lambda_{xy}=(1-\sqrt{1-4g_{xy}})/2$, if
$g_{xy}<1/4$ the ground state (\ref{ground}) is well-defined
because $\lambda_{xy}>0$ is real . Although the interaction
is attractive, the particles in one dimension mutually
exclude
 because  the Jastraw factor in the ground state wave
function forbids particles occupy the same position.
  The ground state energy $E_g$ will be
imaginary if $g_{xy}>\frac{1}4$. This means that the ground
state is not stable when $g_{xy}>\frac{1}4$. The collapse of
the ground state reflects the factorization of the wave
function is false if the attraction between atoms is too
strong. The system is back to a two-dimensional BEC if
$\epsilon_{dd}<1$ because the on-site repulsive is not
renormalized.

For the repulsive coupling constant $g_z$,
$\lambda_\pm=\frac{1\pm\sqrt{1+4g_z}}2$.   For
$-\frac{1}2<\lambda_-<0$, i.e, $g_z<\frac{3}4$, the wave
function is still square integrable. This implies if the
repulsive between particles is not strong enough, the
particles would like to gather together in the real space.
Hence, the long wave length approximation used here is not
valid and one has to be studied in the lattice model. We do
not go this complicated case here.
 For a strong repulsion with
$g_z>\frac{3}4$ , (\ref{ground}) is no longer square
integrable for $\lambda=\lambda_-<-\frac{1}2$. Instead, one
should take $\lambda=\lambda_+>$ 1.5 in (\ref{ground}). This
is the standard solution of the C-S model.

From now on we will focus on the C-S gas, i.e.,
$g_{xy}<\frac{1}4$ and $g_z>\frac{3}4$. For the C-S gas, all
the excitation states may be constructed and may be
expressed by so called hidden-Jack polynomials \cite{hjack}.
We can, in principle, calculate any physical observable.
However, to directly measure these observable in a trapped
gas is difficult. In experiments, a well-established
technique is switching off all trapping potentials and
letting the atom cloud freely expand. After a long enough
time, record the image of the cloud. In this time of flight,
the image of free atom cloud reflects the momentum
distribution of the trapped atoms. Due to the harmonic trap,
the momentum is no longer a good quantum number. Thus, the
calculation from these exact eigen states in the
thermodynamic limit becomes unwieldy. A frequently used
approximation is the local density approximation. The
starting point of this approximation is solving the
homogeneous system in the thermodynamic limit, i.e.,
$\omega_R=0$ and $N$ and $L\to \infty$ while $\bar\rho=N/L$
is fixed. Solving the model in this limit yields to solve
the thermodynamic limit of the Sutherland model with a
period potential \cite{suth}
\begin{eqnarray}
V=\frac{\hbar^2}{2m_R}\sum_{i\ne
j}\frac{\pi^2\lambda(\lambda-1)}{L^2\sin^2[\pi(x_i-x_j)/L]}.
\end{eqnarray}
The quantum states of the Sutherland model are labelled by the
pseudo-momenta $k_i$ which are determined by the Bethe ansatz
equations
$
Lk_i=2\pi I_i+\pi(1-\lambda)\sum_{j<i}{\rm sgn}(k_j-k_i).
$
 Due to the exclusion
between the particles, the ground state is a pseudo-Fermi
sea which is given by $
\{k_j^0\}=\{-\frac{\pi\lambda}L(N-1),-\frac{\pi\lambda}L(N-3),
...,\frac{\pi\lambda}L(N-3),\frac{\pi\lambda}L(N-1)\}$ . The
pseudo-Fermi momentum is $k_F=\frac{\pi\lambda}L(N-1)$.

Notice that the situation in $\lambda<1$ is very different
from $\lambda>1$. For $\lambda>1$, in a physical momentum
interval $\frac{2\pi}L$, there is only either one or zero
{\it one} since the pseudo-momentum interval is
$\frac{2\pi\lambda}L>\frac{2\pi}L$. But for $\lambda<1$, as
we have seen, the number of {\it ones} is larger than one in
an interval $\frac{2\pi}L$. Such a cluster behavior of the
particles stems from the attraction between the particles if
$\lambda<1$. To see this matter clearly, we calculate the
single particle correlation function. In fact, all
correlation functions and their asymptotic forms were
calculated. (See, e.g., \cite{ha, wuy}.) In the
thermodynamic limit, we need only know the asymptotic ones.
For the particle $\Psi_{\lambda}^\dag(x)$ with an exclusion
statistics parameter $\lambda$, the single-particle
correlation function in the $x\to\infty$ limit is given by $
G(x,0;\lambda)\equiv \langle
\Psi_{\lambda}(x,0)\Psi^\dag_{\lambda}(0,0)\rangle \sim
x^{-\lambda}. $ For $\lambda=\lambda_+>1$, the momentum
distribution of the single-particle is well-defined and
continuous near the Fermi momentum $p_F=\hbar k_F/\lambda$,
$ G_{\lambda_+}(p)= G_{\lambda_+}(p_F)-{\rm const.}\cdot
{\rm sgn}(p-p_F)|p-p_F|^{\lambda_+-1}, $ where $p$ is the
physical momentum with interval $\frac{2\pi}L$. Thus, it is
a standard Luttinger liquid \cite{haldl,lutt}.

 However, for $\lambda=\lambda_{xy}<1$, the momentum
distribution of the single-particle is divergent at the
Fermi point $p_F$, i.e., $ G_{\lambda_{xy}}(p)\sim {\rm
sgn}(p-p_F)|p-p_F|^{\lambda_{xy}-1}$. This divergence
implies that there is no single particle propagation. The
instability of this single particle Fermi surface comes from
the attraction between the particles. The competition
between this attraction and the zero-point motion may lead
to a cluster propagating behavior as we have analyzed in the
pseudo-Fermi sea. In the coordinate space, this cluster
behaviors can be seen as follows. Defining a composite
particle operator \cite{haldl} $
\Psi_{n\lambda_{xy}}(x)=\lim_{a\to 0}
[a^{-\frac{1}2n(n-1)}\Psi_{\lambda_{xy}}(x)
\Psi_{\lambda_{xy}}(x+a) ...\Psi_{\lambda_{xy}}(x+(n-1)a)],
$ for an integer $n$, the momentum distribution of this
composite particle near $p_F$ is given by $
G_{n\lambda_{xy}}(p)=G_{n\lambda_{xy}}(p_F)- {\rm
const.}\cdot {\rm sgn}(p-p_F)|p-p_F|^{n\lambda_{xy}-1}. $
The minimal $n$ with $n\lambda_{xy}-1\geq 0$ is
$n=[1/\lambda_{xy}]+1$. However, the composite particle with
$n'=[1/\lambda_{xy}]-1$ will accompany with this
$n=[1/\lambda_{xy}]+1$ particle to insure the pseudo-Fermi
sea is filled up but the former has a divergent distribution
at the Fermi surface. A consistent choice is $n=l$ for
$\lambda_{xy}=s/l$ and thus the exponent is
$l\lambda_{xy}-1=s-1$, e.g., $\lambda_{xy}=2/5$, $s-1=1$.
This reflects the exclusion statistics of the
quasi-particles.

\noindent{\it Time of flight and local density
approximation} The time of flight experiment directly
measure the momentum distribution in the trapped gas. In the
$y$-direction along the cigar-like BEC, the dynamics is
controlled by Bogoliubov quasiparticles excitation while in
the $x$-direction, the dynamics is dominated by the
Luttinger liquid described above. We hope this Luttinger
liquid behavior may be observed by measuring the slope of
the density profile at the Fermi surface in the
$x$-direction. The density-density correlation function
$\langle\hat{\rho}(x,0)\hat{\rho}(0,0)\rangle\sim x^{-2}$
corresponds to the shot noise in the time of flight image
\cite{altman}. To experimentally observe this Luttinger
liquid requires a very high resolving rate because there is
background $LN_0$ atoms' image.

 For $\omega_R\ne 0$, one may use the local density
approximation. Defining the local Fermi momentum $
 p_F(x)=\sqrt{p_F^2-m^2_R\omega_R^2x^2/\lambda^2},
$
 the density distribution in the Thomas-Fermi approximation, which is
 valid in
 the thermodynamic limit, is defined by \cite{butt}
\begin{eqnarray}
 n(p)&=&\frac{1}{2\pi}\int
 ^{p_F}_{p_F(L)}dP \frac{\lambda}{m_R\omega_R}\frac{ G_{\lambda}(p,P)P}{\sqrt{p_F^2-P^2}
}\nonumber\\ &\approx&\frac{\lambda^2}{4\pi\hbar
m_R^2\omega_R^2}\frac{\bar p_F}{\sqrt{p_F^2-\bar
p_F^2}}G_\lambda(p,\bar p_F)
\end{eqnarray}
 where $p_F(L)=\sqrt{p_F^2-m_R\hbar\omega_R/\lambda}$ and $\bar p_F$
 is in between $p_F$ and $p_F(L)$. Unlike
the short range interaction Bose gas, the long range
interaction gas in one-dimension is more like the degenerate
Fermi gas: Despite the spatial anisotropy of the trap, the
momentum distribution is uniform in the $k_x$-direction
\cite{butt}.

\noindent{\it Conclusions} We have shown the possibility to
realize the C-S gas in a cold atom system with dipole-dipole
interaction. The ground state behaviors of the system in
various parameters were discussed. The experimental
implications to explore the properties of these ground
states and low-lying excitations are suggested.

 The author thanks Su Yi and Li You for useful discussions. This work was supported in part by Chinese
National Natural Science Foundation.


\vspace{-0.5cm}




\begin{references}

\vspace{-0.5cm}

\bibitem{kohl} T. St\"oferle et al, Phys. Rev. Lett. {\bf 92}, 130403(2004).
\bibitem{TG} L. Tonks, Phys. Rev. {\bf 50}, 955(1936); M.
Girardeau, J. Math. Phys. {\bf 1}, 516 (1960).
\bibitem{TGE} T. Kinoshita et al, Sciences {\bf
305}, 1125 (2004); B. Paredes et al, Nature {\bf 429},277 (2004).
\bibitem{ca} F. Calogero, J. Math. Phys. {\bf 10}, 2191 and 2197 (1969).
\bibitem{suth} B. Sutherland, J. Math. Phys. {\bf 12}, 246 (1971); Phys. Rev.
A {\bf 4}, 2019 (1971); {\bf 5}, 1372 (1972).
\bibitem{jack} H. Jack, Proc. Roy. Soc. Edinburgh Sect. A {\bf
69},1 (1969-1970).
\bibitem{ha} See e.g., Z. N. C. Ha, Nucl. Phys. B {\bf 435}, 604
(1995).
\bibitem{hald1} F. D. M. Haldane, Phys. Rev.
Lett. {\bf 67}, 937 (1991).
\bibitem{wu} Y. S. Wu, Phys. Rev. Lett.
{\bf73}, 922 (1994).
\bibitem{wub} D. Bernard and Y. S. Wu,
in Proc. 6th Nankai Workshop, eds. M. L. Ge and Y. S. Wu, World
Scientific (1995).
\bibitem{haldl} F. D. M. Haldane, J. Phys. C {\bf 14}, 2585
(1981).
\bibitem{lutt}
A. Luther and I. Peschel, Phys. Rev. B {\bf 9}, 2911 (1974); E. H.
Lieb and D. C. Mattis, J. Math. Phys. {\bf 6}, 304(1965); J. M.
Luttinger, J. Math. Phys. {\bf 4}, (1963); S. Tomonaga, Prog.
Theor. Phys. {\bf 5}, 544 (1950).
\bibitem{wuy} Y. S. Wu and Y. Yu, Phys. Rev. Lett. {\bf 75}, 890
(1995); Y. S. Wu, Y. Yu and H. X. Yang, Nucl. Phys. B {\bf 604},
551 (2001).
\bibitem{KY} N. Kawakami and S. K. Yang, Phys. Rev. Lett. {\bf 67}, 2493
(1990).
\bibitem{cft} R. Dijkgraaf et al,
Comm. Math. Phys. {\bf 115}, 649 (1988).
\bibitem{sla}B. D. Simons, P. A. Lee, and B. L. Altshuler, Phys. Rev. Lett.
70, 4122 (1993).
\bibitem{yu} Y. Yu and Z. Y.  Zhu, Comm. Theor. Phys. {\bf 29},
 351 (1998); Y. Yu, W. J. Zheng and
Z. Y. Zhu, Phys. Rev. B {\bf 56}, 13279 (1997); W. J.  Zheng and
Y. Yu,  Phys. Rev. Lett. {\bf 79}, 3242 (1997).
\bibitem{csmb} See e.g., J. F. van Diejen and L. Vinet (Eds.),
{\it Calogero-Moser-Sutherland Models: CRM Series in Mathematical
Physics 2000 XXV}, (Springer, Berlin, Heidelberg, New York 2000)
and references therein.
\bibitem{dpdp} A. Griesmaier, J. Werner, S. Hensler, J. Stuhler,
 and T. Pfau, Phys. Rev. Lett. {\bf 94}, 160401(2005).
\bibitem{dpt1} S. Yi and L. You, Phys.
Rev. A {\bf 61}, 041604(R) (2000); L. Santos et al, Phys. Rev.
Lett. {\bf 85}, 1791 (2000); K. Goral et al., Phys. Rev. A {\bf
61}, 051601(R) (2000);J.- P. Martikainen et al., Phys. Rev. A {\bf
64}, 037601 (2001); D. H. J. O'Dell et al, Phys. Rev. Lett. {\bf
92}, 250401 (2004).
\bibitem{dpt2}   S. Yi and L. You, Phys. Rev.
A {\bf 63}, 053607 (2001); S. Giovanazzi et al, Phys. Rev. Lett.
{\bf 88}, 130402 (2002);   D. DeMille, Phys. Rev. Lett. {\bf 88},
067901 (2002); P. M. Lushnikov, Phys. Rev. A {\bf 66}, 051601(R)
(2002); U. R. Fischer, Phys. Rev. Lett. {\bf 89}, 280402 (2002);
B. Damski et al., Phys. Rev. Lett. {\bf 90}, 110401 (2003); S. Yi
and L. You, Phys. Rev. A {\bf 67}, 045601 (2003); A. Derevianko,
Phys. Rev. A {\bf 67}, 033607 (2003);  A. V. Avdeenkov et al,
Phys. Rev. A {\bf 69}, 012710 (2004); S. Yi and L. You, Phys. Rev.
Lett. {\bf 92}, 193201 (2004).

\bibitem{van} D. van Oosten, P. van der Straten, and H. T. C.
Stoof, Phys. Rev. A {\bf 67}, 033606 (2003).

\bibitem{li} The renormalization of $t$ has been considered by J. B. Li, Y. Yu, A. M. Dudarev and Q. Niu,
New J. Phys  {\bf 8}, 154 (2006).

\bibitem{epsilon} A. Griesmaier, J. Stuhler, T. Koch,
M. Fattori, T. Pfau, and S. Giovanazzi, E-preprint,
cond-mat/0608171.


\bibitem{hjack} See e.g., M. Wadati and H. Ujino in
Ref.\cite{csmb}.

\bibitem{altman} E. Altman et al, Phys. Rev.
A {\bf 70}, 013603 (2004).

\bibitem{butt} D. A. Butts and D. S. Rokhsar, Phys. Rev. A {\bf
55}, 4346 (1997).
\end{references}
\end{document}